\documentclass[final,3p,times,twocolumn]{elsarticle}

\usepackage{graphicx}
\usepackage{hyperref}
\usepackage{atlasphysics}

\usepackage{multirow}
\usepackage{amsmath}
\usepackage{subfigure}
\usepackage{rotating}
\usepackage{wasysym}

\newcommand{\atlasnote}[1]{\def\@atlasnote{#1}}

\newcommand{\ggF}{\ensuremath{gg\to H}}
\newcommand{\VBF}{\ensuremath{qq'\to qq'H}}
\newcommand{\VH}{\ensuremath{q\bar{q}\to WH/ZH}}
\newcommand{\ttH}{\ensuremath{q\bar{q}/gg \to t\bar{t}H}}

\newcommand{\mll}{$m_{\ell\ell}$}

\newcommand{\hgg}{\ensuremath{H \rightarrow \gamma\gamma}}

\newcommand{\hWWlnln}{\ensuremath{H \rightarrow WW^{(*)} \rightarrow \ell^+\nu\ell'^-\overline{\nu}}}

\newcommand{\hWWlnqq}{\ensuremath{H \rightarrow WW \rightarrow \ell\nu q\overline{q}'}}

\newcommand{\hZZllnn}{\ensuremath{H \rightarrow ZZ\rightarrow \ell^+\ell^-\nu\overline{\nu}}}
\newcommand{\hZZllqq}{\ensuremath{H \rightarrow ZZ\rightarrow \ell^+\ell^- q\overline{q}}}
\newcommand{\hZZllll}{\ensuremath{H \rightarrow ZZ^{(*)}\rightarrow \ell^+\ell^-\ell'^+\ell'^-}}

\newcommand{\mh}{\ensuremath{m_{H}}}

\newcommand{\lowerExp}{124~GeV}
\newcommand{\lowerExpNoGeV}{124}
\newcommand{\upperExp}{519~GeV}
\newcommand{\lowerlowerObs}{131~GeV}
\newcommand{\lowerlowerObsNoGeV}{131}
\newcommand{\upperlowerObs}{238~GeV}

\newcommand{\lowerIsland}{112.9~GeV}
\newcommand{\lowerIslandNoGeV}{112.9}
\newcommand{\upperIsland}{115.5~GeV}
\newcommand{\lowerupperObs}{251~GeV}
\newcommand{\lowerupperObsNoGeV}{251}
\newcommand{\upperupperObs}{466~GeV}
\newcommand{\significance}{$3.6\sigma$}
\newcommand{\significanceESS}{$3.5\sigma$}

\newcommand{\ZZbkg}{\ensuremath{ZZ^{(*)}}}

\newcommand{\infb}{fb$^{-1}$}

\usepackage{amssymb}
\usepackage{lineno}

\usepackage{chngpage}

\biboptions{numbers,square,comma,sort&compress}

\begin{document}

\begin{frontmatter}

\title{\vspace{-0.5cm} \flushright{\normalsize{CERN-PH-EP-2012-019}} \\ \vspace{3.cm} \begin{center} Combined search for the Standard Model Higgs boson using up to
  4.9~\infb\ \\ of $pp$ collision data at $\sqrt{s}=7$~TeV with the
  ATLAS detector at the LHC \end{center}}

\begin{abstract}
  A combined search for the Standard Model Higgs boson with the ATLAS
  experiment at the LHC using datasets corresponding to integrated
  luminosities from 1.04~\infb\ to 4.9~\infb\ of $pp$ collisions
  collected at $\sqrt{s}=7$~TeV is presented. The Higgs boson mass
  ranges \lowerIslandNoGeV --\upperIsland, \lowerlowerObsNoGeV
  --\upperlowerObs\ and \lowerupperObsNoGeV --\upperupperObs\ are
  excluded at the 95\%\ confidence level (CL), while the range
  \lowerExpNoGeV --\upperExp\ is expected to be excluded in the
  absence of a signal. An excess of events is observed around
  \mh$\sim$126~GeV with a local significance of 3.5 standard
  deviations ($\sigma$).  The local significance of \hgg, \hZZllll\
  and \hWWlnln, the three most sensitive channels in this mass range,
  are 2.8$\sigma$, 2.1$\sigma$ and 1.4$\sigma$, respectively. The
  global probability for the background to produce such a fluctuation
  anywhere in the explored Higgs boson mass range 110--600~GeV is
  estimated to be $\sim$1.4\% or, equivalently, $2.2\sigma$.
\end{abstract}

\journal{Physics Letters B}
\author{\begin{center} The ATLAS Collaboration \end{center}}

\end{frontmatter}

\hyphenation{ATLAS}

\section{Introduction}

The discovery of the mechanism for electroweak symmetry breaking
(EWSB) is a major goal of the physics programme at the Large Hadron
Collider (LHC). In the Standard Model (SM), EWSB is achieved by
invoking the Higgs mechanism, which requires the existence of the
Higgs
boson~\cite{Englert:1964et,Higgs:1964ia,Higgs:1964pj,Guralnik:1964eu,
  Higgs:1966ev,Kibble:1967sv}. In the SM, the Higgs boson mass, \mH,
is a priori unknown. However, for a given \mH\ hypothesis, the
production cross sections and branching fractions of each decay mode
are predicted, which enables a combined search with data from several
decay channels.

Direct searches at the CERN LEP $e^+e^{-}$ collider excluded the
production of a SM Higgs boson with mass below 114.4~GeV at
the 95\%~CL~\cite{LEP}. The combined searches at the Fermilab Tevatron
$p\overline{p}$̄ collider have excluded the production of a Higgs
boson with mass between 156~GeV and 177~GeV at
the 95\%~CL~\cite{tevatronHiggs}.

In 2011, the LHC delivered to ATLAS an integrated luminosity of
5.6~\infb\ of $pp$ collisions at 7~TeV centre-of-mass energy. The
ATLAS experiment collected and analysed an integrated luminosity
corresponding to up to 4.9~\infb\ of data fulfilling all the data
quality requirements to search for the SM Higgs boson. In
this Letter a combined search using six distinct channels, covering
the mass range 110~GeV to 600~GeV, is presented.  The Higgs boson is
produced primarily through the gluon fusion process and the following
decay modes are considered: \hgg, \hZZllll, \hZZllqq, \hZZllnn,
\hWWlnln, and \hWWlnqq, where $\ell$ denotes an electron or a muon.
New limits on SM Higgs boson production are established and
the significance of an excess of events observed in the low mass
region around \mH=126~GeV is quantified.

\section{Search Channels}

All search analyses are described in their respective
references~\cite{ggPaper,4lPaper,lvlvPaper,lvqqPaper,llvvPaper-1fb,llqqpaper-1fb}
and therefore only the main features relevant to the statistical
combination of the various channels are summarised here. Two channels,
the \hZZllqq\ and \hZZllnn, have been updated to a data sample
corresponding to an integrated luminosity larger than that used in the
previously published results and are described in more detail.

\begin{figure}[tb]
  \begin{center}
    \includegraphics[width=.235\textwidth]{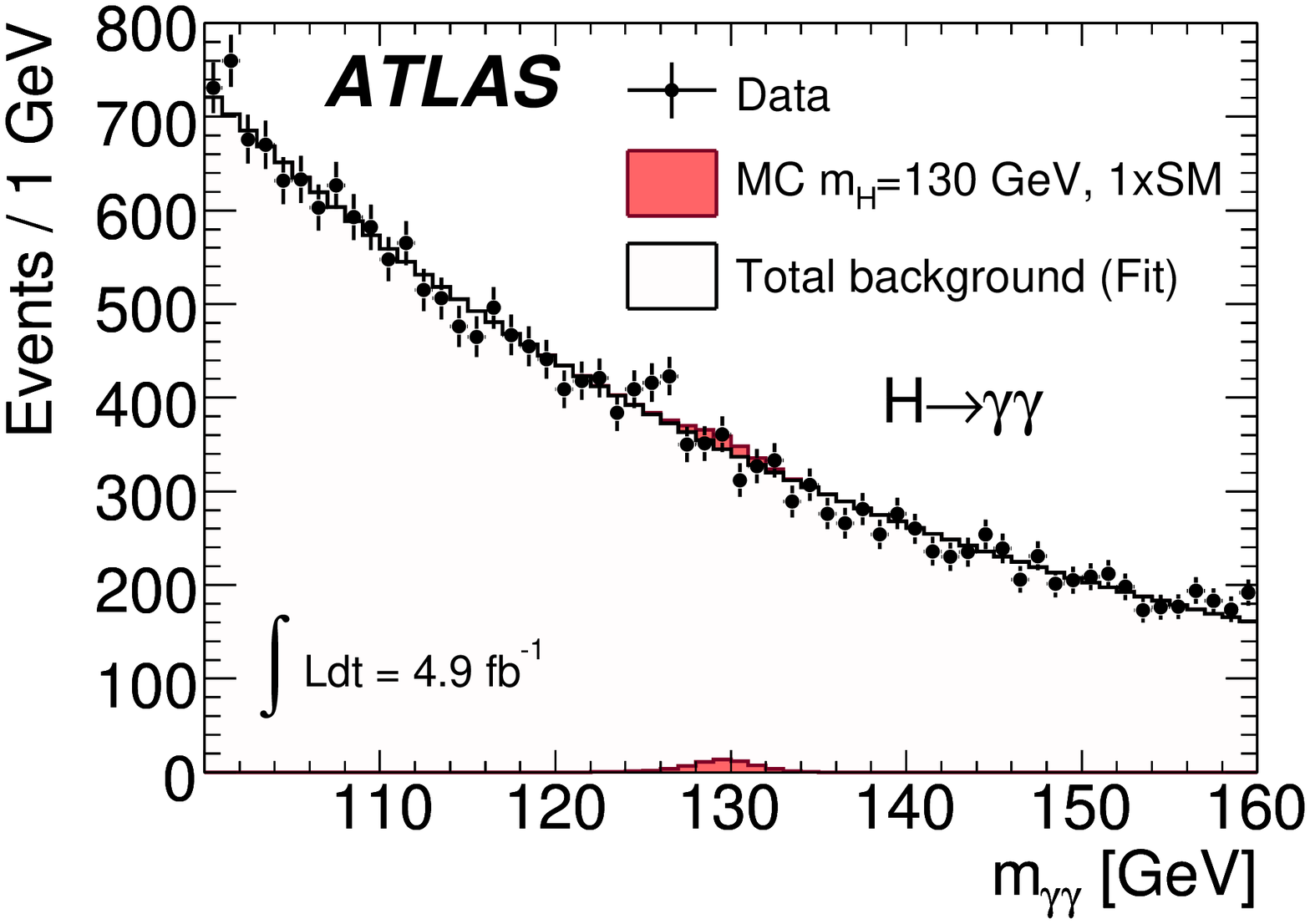} \hspace{-0.2cm}
    \includegraphics[width=.235\textwidth]{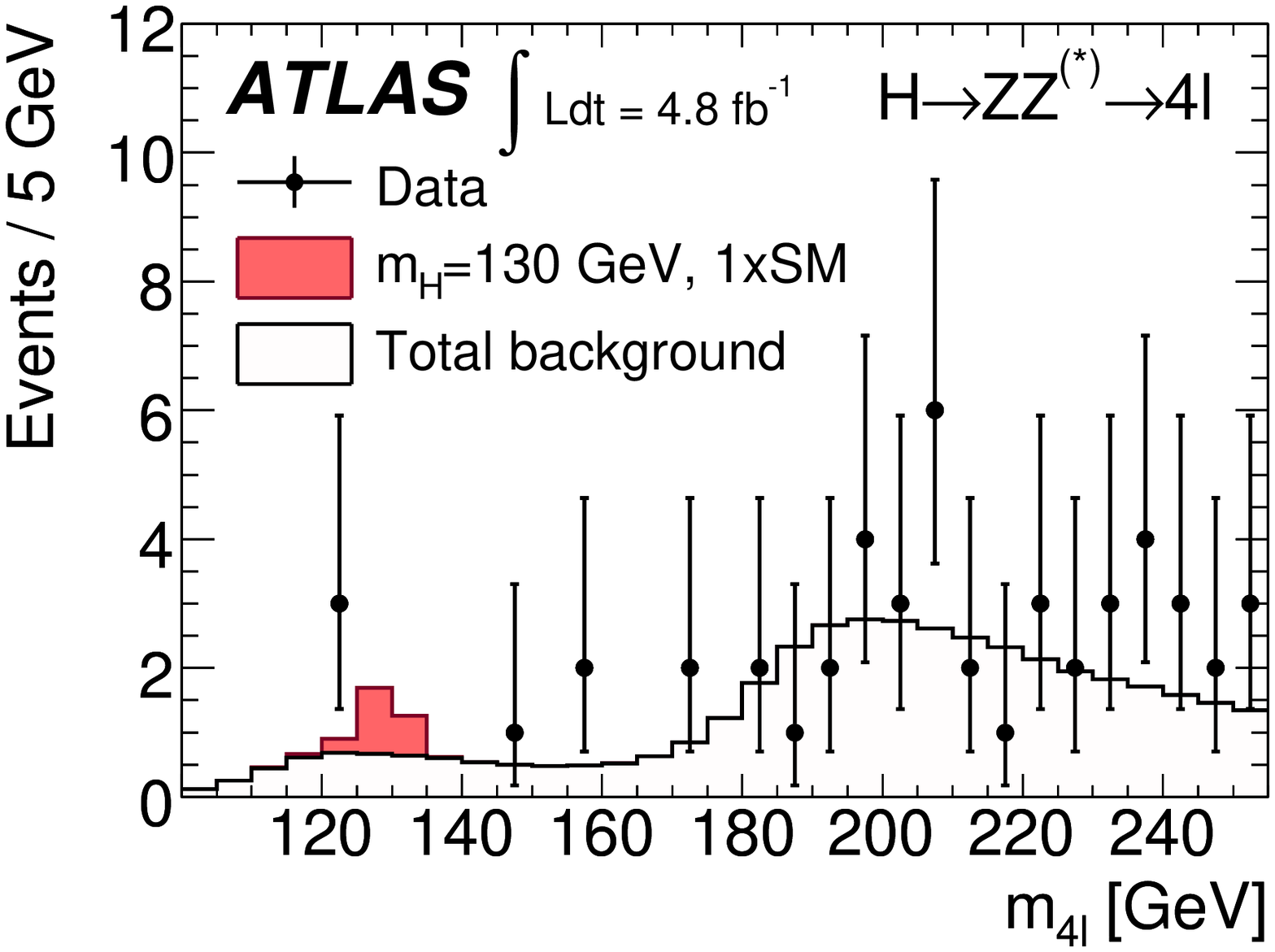} \hspace{-0.2cm} \\
     \vspace{-0.3cm}
    \hspace{.02\textwidth}  {\footnotesize (a)}  \hspace{.21\textwidth}  {\footnotesize (b)} \\
     \vspace{0.15cm}
    \includegraphics[width=.235\textwidth]{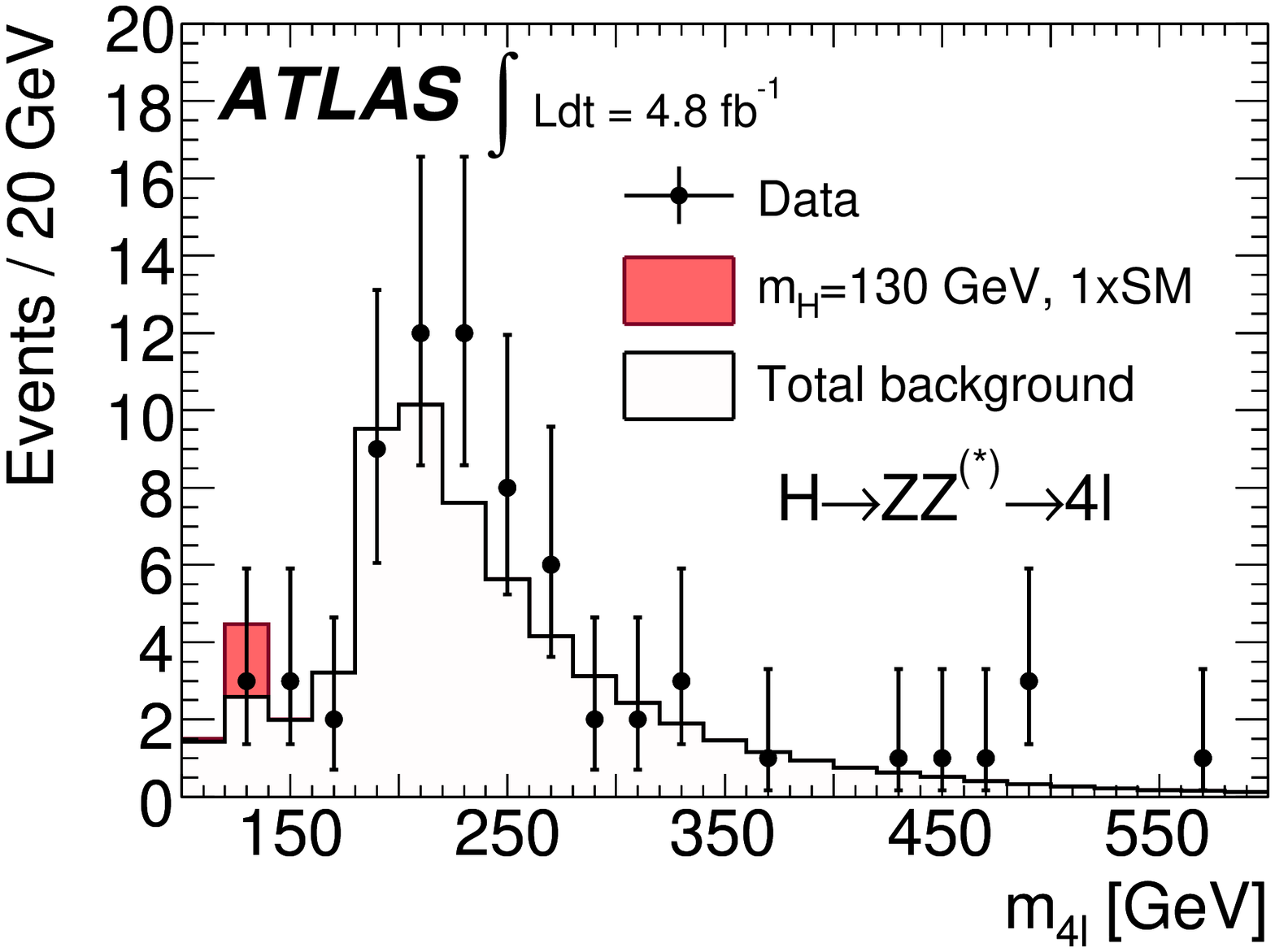} \hspace{-0.2cm}
    \includegraphics[width=.235\textwidth]{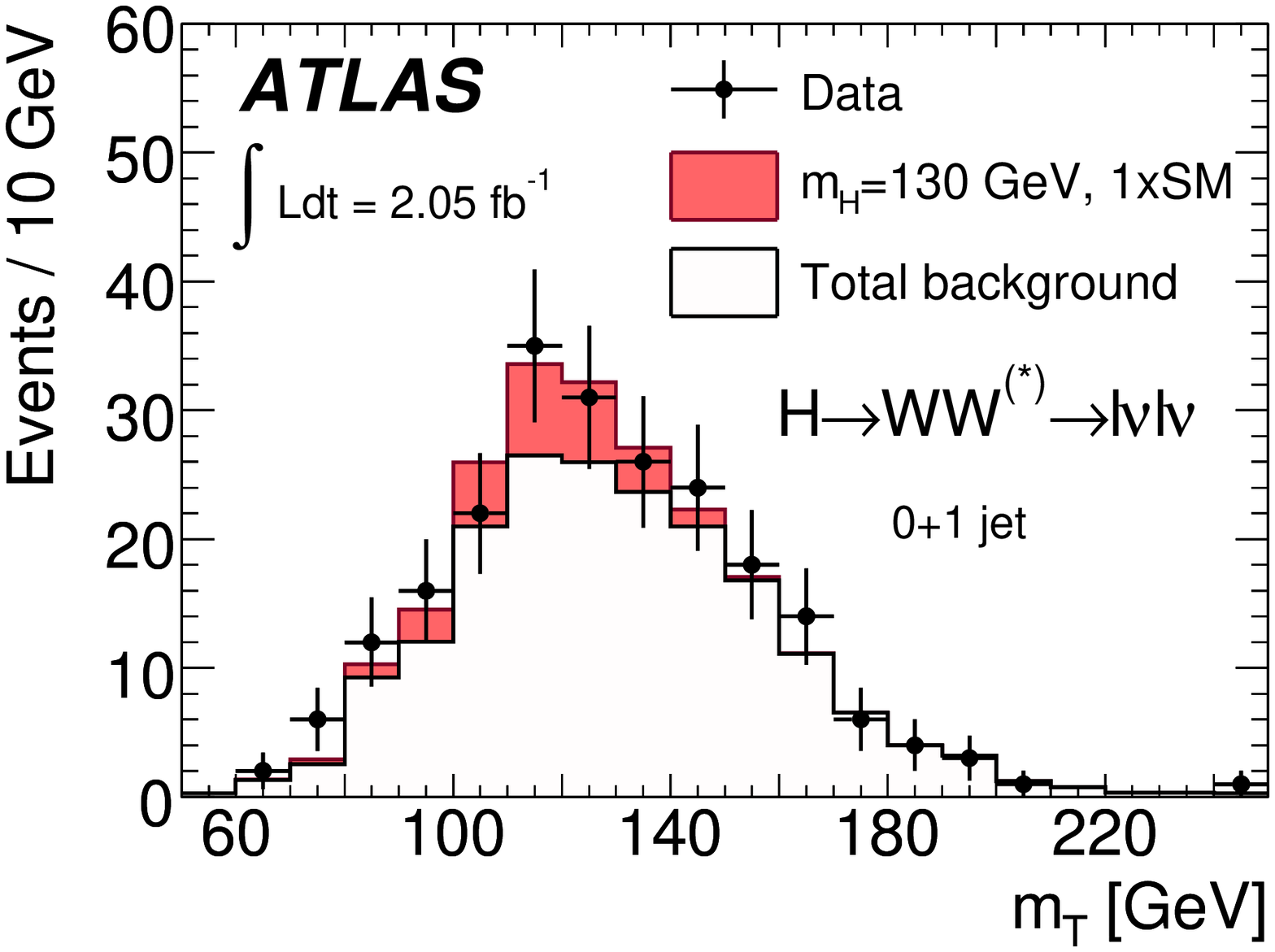} \\
     \vspace{-0.3cm}
    \hspace{.02\textwidth}  {\footnotesize (c)}  \hspace{.21\textwidth}  {\footnotesize (d)} \\
     \vspace{-0.6cm}
  \end{center}
  \caption{Distributions of the reconstructed invariant or transverse
    mass for the selected candidate events and for the total
    background and signal (\mH=130~GeV) expected in the \hgg\ (a), the
    \hZZllll\ in the low mass region (b) and the entire mass range
    (c), and the \hWWlnln\ (d) channels.}
  \label{fig:lowmass}
\end{figure}

The \hgg\ search is carried out for \mH\ hypotheses between 110~\GeV\
and 150~\GeV\ and uses an integrated luminosity of
4.9~\infb~\cite{ggPaper}.  The analysis in this channel separates
events into nine independent categories of varying sensitivity. The
categorisation is based on the direction of each photon and whether it
was reconstructed as a converted or unconverted photon, together with
the momentum component of the diphoton system transverse to the thrust
axis. The diphoton invariant mass $m_{\gamma\gamma}$ is used as a
discriminating variable to distinguish signal and background, to take
advantage of the mass resolution of approximately 1.4\%\ for
\mh$\sim$120~GeV. The distribution of $m_{\gamma\gamma}$ in the data
is fit to a smooth function to estimate the background. The inclusive
invariant mass distribution of the observed candidates, summing over
all categories, is shown in Fig.~\ref{fig:lowmass}(a).

The search in the \hZZllll\ channel is performed for \mH\ hypotheses
in the full 110~\GeV\ to 600~\GeV\ mass range using data corresponding
to an integrated luminosity of 4.8~\infb~\cite{4lPaper}.  The main
irreducible \ZZbkg\ background is estimated using Monte Carlo
simulation. The reducible $Z$+jets background, which has an impact
mostly for low four-lepton invariant masses, is estimated from control
regions in the data. The top-quark background normalisation is
validated in a control sample of events with an opposite sign
electron-muon pair with an invariant mass consistent with that of the
$Z$ boson and two leptons of the same flavour. The events are
categorised according to the lepton flavour combinations. The mass
resolutions are approximately 1.5\%\ in the four-muon channel and 2\%\
in the four-electron channel for \mh$\sim$120~GeV. The four-lepton
invariant mass is used as a discriminating variable. Its distribution
for events selected after all cuts is displayed in
Fig.~\ref{fig:lowmass}(b) for the low mass range and
Fig.~\ref{fig:lowmass}(c) for the full mass range.

\begin{figure}[tb]
  \begin{center}
    \includegraphics[width=.235\textwidth]{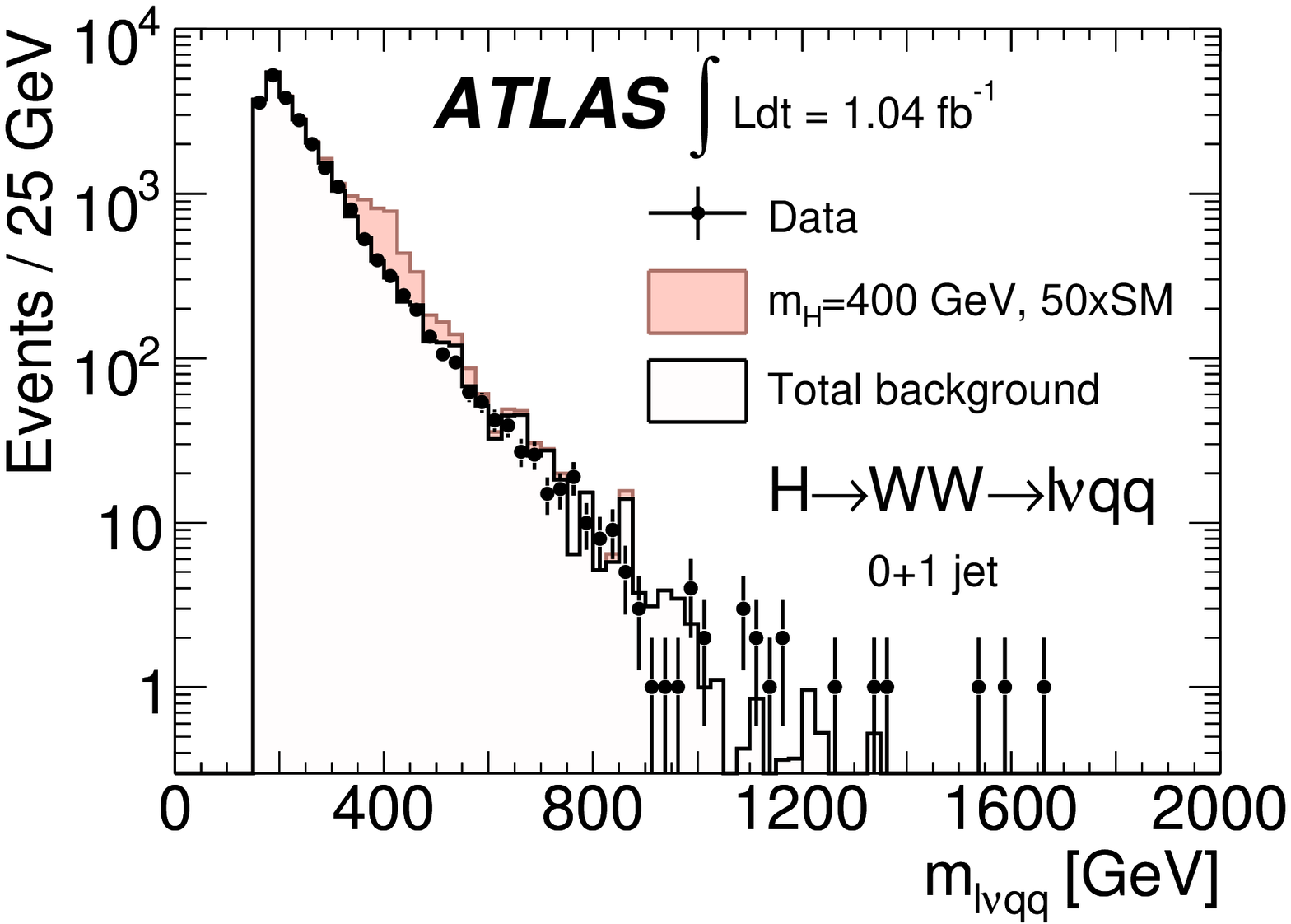} \hspace{-0.2cm}
    \includegraphics[width=.235\textwidth]{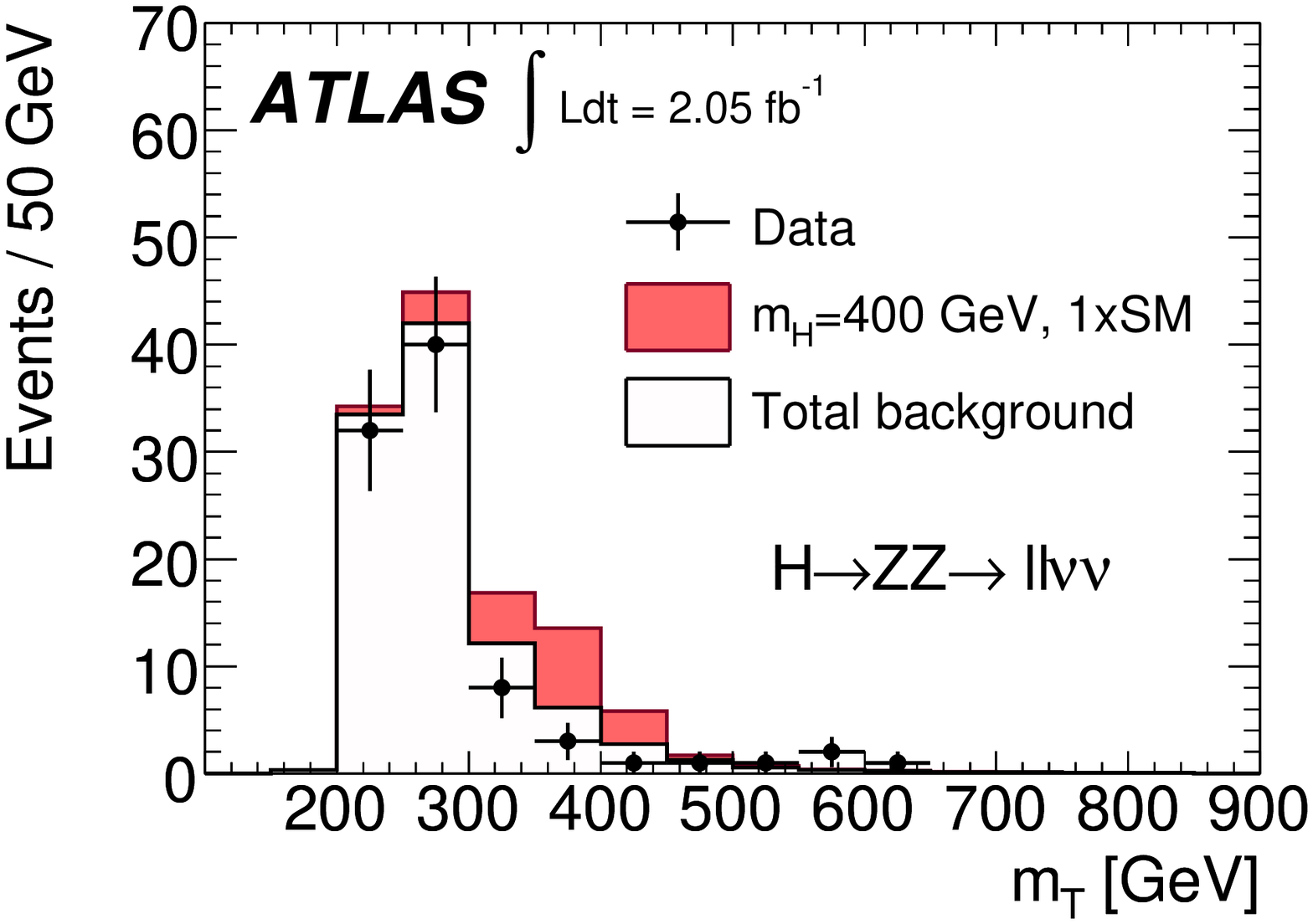}  \\
     \vspace{-0.3cm}
    \hspace{.02\textwidth}  {\footnotesize (a)}  \hspace{.21\textwidth}  {\footnotesize (b)} \\
     \vspace{0.15cm}
    \includegraphics[width=.235\textwidth]{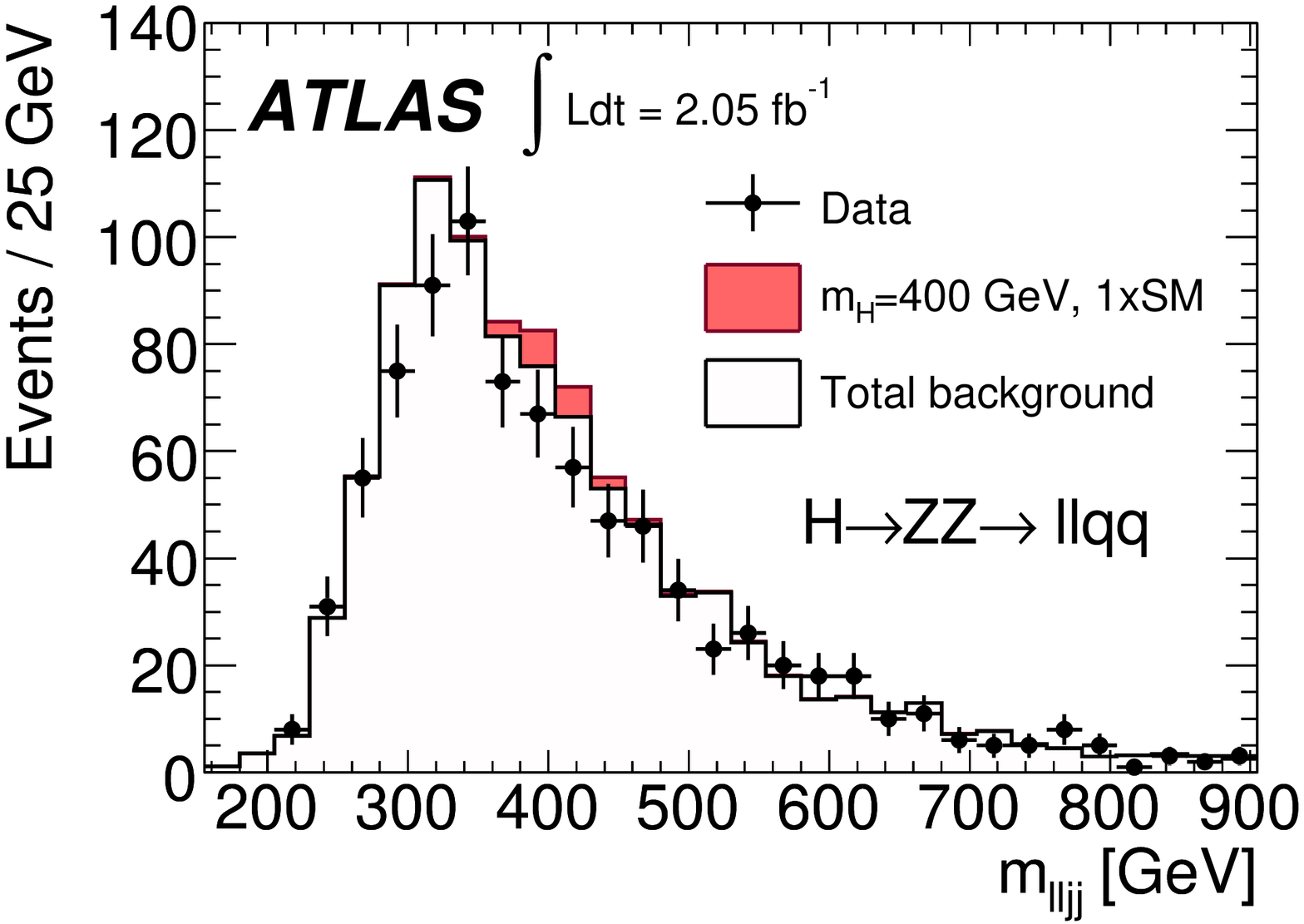} \hspace{-0.2cm}
    \includegraphics[width=.235\textwidth]{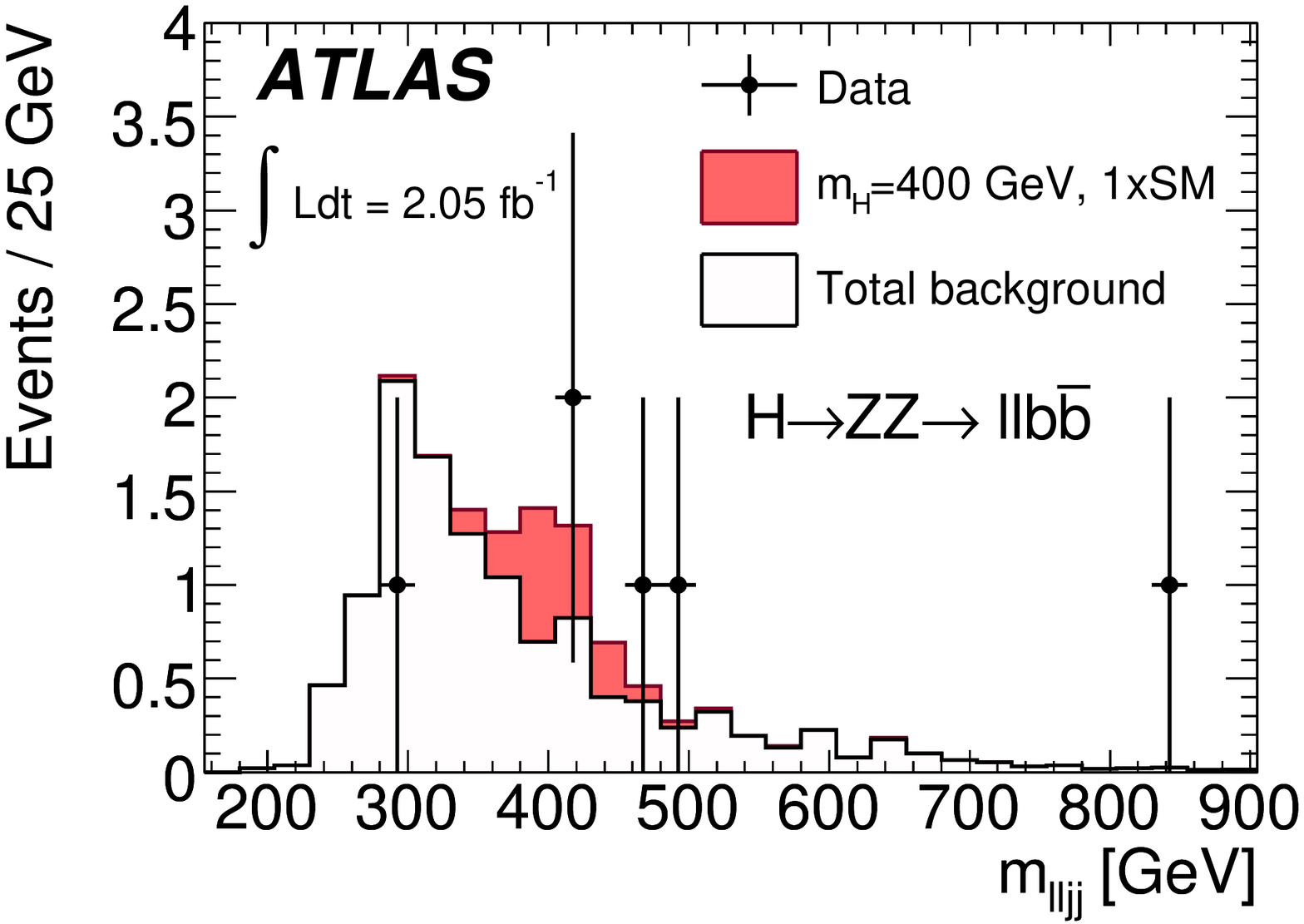} \\
     \vspace{-0.3cm}
    \hspace{.02\textwidth}  {\footnotesize (c)}  \hspace{.21\textwidth}  {\footnotesize (d)} \\
     \vspace{-0.6cm}
  \end{center}
  \caption{Distributions of the reconstructed invariant or transverse
    mass, in analyses relevant for the search of the Higgs boson at
    high mass, for selected candidate events, the total background and
    the signal (\mH=400~GeV) expected for the given value of \mh\ in
    the \hWWlnqq\ channel (a), the \hZZllnn\ channel (b) and the \hZZllqq\
    channel for events selected in the $b$-jet untagged (c) and
    the tagged (d) categories. The signal distribution is
    displayed in a lighter red colour in the \hWWlnqq\ channel where
    it has been scaled up by a factor 50.}
  \label{fig:highmass}
\end{figure}

The \hWWlnln\ search is performed as an event counting analysis for
\mH\ hypotheses between 110~\GeV\ and 300~\GeV, using an integrated
luminosity of 2.05~\infb~\cite{lvlvPaper}.  The main background
contribution, from non-resonant $WW$ production, is estimated from the
data using control regions based on the dilepton invariant mass
\mll. The analysis is separated into 0-jet and 1-jet categories as
well as according to lepton flavour. In the 1-jet category, a $b$-jet
veto is applied to reject events from top-quark production. The
relative fractions of the background contributions expected in the
signal and control regions are taken from Monte Carlo simulation.  The
transverse mass distribution of events for both jet categories is
displayed in Fig.~\ref{fig:lowmass}(d).

The \hWWlnqq\ analysis covers \mH\ hypotheses in the 240~\GeV\ to
600~\GeV\ range and is carried out using data corresponding to an
integrated luminosity of 1.04~\infb~\cite{lvqqPaper}. This channel is
also separated according to lepton flavour and into 0-jet and 1-jet
categories, where the number of jets refers to those in addition to
the jets selected as originating from the $W$-boson decay. Events with
at least one $b$-tagged jet are rejected to reduce backgrounds from
top-quark production.  The $\ell \nu q\overline{q}'$ mass is
reconstructed using a constraint to the $\ell\nu$ system to $W$-boson
mass. It is used as a discriminating variable and its distribution is
illustrated in Fig.~\ref{fig:highmass}(a).

The search in the \hZZllnn\ channel is performed in the 200~\GeV\ to
600~\GeV\ range of \mH. The analysis described in
Ref.~\cite{llvvPaper-1fb} is based on data corresponding to an
integrated luminosity of 1.04~\infb. It has been updated using a
dataset corresponding to an integrated luminosity of 2.05~\infb. The
main change in the event selection is the use of an improved
$b$-tagging algorithm~\cite{BTagging} to veto events with jets likely
to have originated from $b$-quarks. The analysis is tuned for two
search regions with \mH\ hypotheses above and below 280~GeV and
separated into lepton flavour categories. The $\ell^+\ell^{-}$ pair
invariant mass is required to be within 15~GeV of the $Z$-boson
mass. The reverse requirement is applied to same flavor leptons in the
\hWWlnln\ channel to avoid overlaps. The transverse
mass is used as a discriminating variable. Its distribution is shown
in Fig.~\ref{fig:highmass}(b) for the high \mH\ search.  In total 175
events are selected in the low \mH\ search and $192 \pm 23$ are
expected from the background. Similarly, the high-mass search selects
89 events, while $100 \pm 11$ are expected from the background. The
expected number of signal events in the low mass search for
\mH=200~GeV is $9.9\pm 1.8$ and $19.6 \pm 3.4$ for the high-mass
selection for \mH=400~GeV.

The analysis of the \hZZllqq\ channel, carried out in the \mH\ range
from 200~\GeV\ to 600~\GeV\ using data corresponding to an integrated
luminosity of 1.04~\infb, is described in
Ref.~\cite{llqqpaper-1fb}. It has been updated using a dataset
corresponding to an integrated luminosity of 2.05~\infb, taking
advantage of the improved $b$-tagging algorithm and of the larger
sample of data to better constrain systematic uncertainties on the
background yield. The analysis is separated into search regions above
and below \mH=300~GeV, where the event selections are independently
optimised.  The dominant background arises from $Z$+jets production,
which is normalised from data using the sidebands of the dilepton
invariant mass distribution. To profit from the sizable branching
fraction of the $Z$ decaying into a pair of $b$-quarks in the signal,
the analysis is divided into two categories, the first containing
events where the two jets are $b$-tagged and the second with events
with fewer than two $b$-tags. Using the $Z$ boson mass constraint
improves the mass resolution of the $\ell{\ell}qq$ system by
approximately 10\%. The number of events selected in the data with the
low \mH\ (high \mH) untagged search is 21000 (851) where $21370
\pm 310$ ($920 \pm 100$) are expected from the background, and $67 \pm
11 $ ($21.1 \pm 0.8$) from a signal with \mH=200~GeV
(\mH=400~GeV). For the tagged search, the number of observed
events in the data with the low \mH\ (high \mH) selection is 145 (6),
in reasonable agreement with the $165 \pm 22 $ ($11.6 \pm 1.9$)
expected from the background, while $4.4 \pm 1.2$ ($2.1 \pm 3.4$) are
expected from a signal with \mH=200~GeV (\mH=400~GeV). The invariant
mass is used as the discriminating variable and its distribution is
shown in Figs.~\ref{fig:highmass}(c) and \ref{fig:highmass}(d) for the
two categories.

\section{Systematic Uncertainties}
\label{sec:syst}

The sources of systematic uncertainties, and their effects on the
signal and background yields and shapes in each individual channel,
are described in detail in
Refs.~\cite{ggPaper,lvlvPaper,lvqqPaper,4lPaper,llvvPaper-1fb,llqqpaper-1fb}. In
the combination, systematic uncertainties are considered either as
fully correlated or uncorrelated.  Partial correlations are treated by
separating a given source into correlated and uncorrelated
components.  The effect of each uncertainty is estimated independently
for each channel. The dominant correlated systematic uncertainties are
those on the measurement of the integrated luminosity and on the
theoretical predictions of the signal production cross sections and
decay branching fractions, as well as those related to detector
response that impact the analyses through the reconstruction of
electrons, photons, muon, jets, magnitude of the missing transverse
momentum ($E_{\rm T}^{\rm miss}$ ) and $b$-tagging.

The uncertainty on the integrated luminosity is considered as fully
correlated among channels and ranges from 3.7\%\ to 3.9\%\ depending
on the data-taking period of the samples used in each specific
channel~\cite{Luminosity, LuminosityCONF}. The uncertainty is larger
for the last part of the 2011 data due to an increase in the average
number of proton-proton interactions occurring in the same bunch
crossing (pileup events).

The Higgs boson production cross sections are computed up to
Next-to-Next-to-Leading Order
(NNLO)~\cite{Djouadi:1991tka,Dawson:1990zj,Spira:1995rr,Harlander:2002wh,Anastasiou:2002yz,Ravindran:2003um}
in QCD for the gluon fusion (\ggF) process, including soft-gluon
resummations up to Next-to-Next-to-Leading Log
(NNLL)~\cite{Catani:2003zt,deFlorian:2011xf} and Next-to-Leading Order (NLO) electroweak
(EW) corrections~\cite{Aglietti:2004nj,Actis:2008ug}.  These results
are compiled in
Refs.~\cite{Anastasiou:2008tj,deFlorian:2009hc,Baglio:2010ae}.  The
cross section for the vector-boson fusion (\VBF) process is estimated
at NLO~\cite{Ciccolini:2007jr,Ciccolini:2007ec,Arnold:2008rz} and
approximate NNLO QCD~\cite{Bolzoni:2010xr}.
The associated $WH$/$ZH$ production processes (\VH) are computed at
NLO~\cite{Han:1991ia,Ciccolini:2003jy} and
NNLO~\cite{Brein:2003wg}.
The associated production with a \ttbar\ pair (\ttH) is estimated at
NLO~\cite{Beenakker:2001rj,Beenakker:2002nc,Dawson:2002tg,Dawson:2003zu}.
The Higgs boson production cross sections, decay branching
ratios~\cite{Djouadi:1997yw,Bredenstein:2006rh,Bredenstein:2006ha,Actis:2008ts}
and their related uncertainties are compiled in
Ref.~\cite{LHCHiggsCrossSectionWorkingGroup:2011ti}.  The QCD scale
uncertainties for \mH=120~GeV amount to $^{+12}_{-8}$\%\ for the \ggF\
process, $\pm1$\%\ for the \VBF\ and associated $WH/ZH$ processes, and
$^{+3}_{-9}$\%\ for the \ttH\ process. The uncertainties related to
the parton distribution functions (PDF) for low \mH\ hypotheses
typically amount to $\pm 8$\%\ for the predominantly gluon-initiated
processes \ggF\ and \ttH, and $\pm 4$\%\ for the predominantly
quark-initiated \VBF\ and $WH/ZH$
processes~\cite{Botje:2011sn,Lai:2010vv,Martin:2009iq,Ball:2011mu}.
The theoretical uncertainty associated with the exclusive Higgs boson
production process with one additional jet in the \hWWlnln\ channel
amounts to $\pm 20$\%\ and is treated according to the prescription of
Refs.~\cite{PhysRevD.85.034011,LHC-HCG,Dittmaier:2012vm}.  Additional
theoretical uncertainty on the signal normalisation, to account for
effects related to off-shell Higgs boson production and interference
with other SM processes, is assigned at high Higgs boson masses
(\mH~$\gtrsim 300$~GeV) as
150\%$\times$(\mH/TeV)$^3$~\cite{lhcCombination,Passarino:2010qk,Anastasiou:2011pi,Dittmaier:2012vm}.

The detector-related sources of systematic uncertainty are modelled
using the following classification: trigger and identification
efficiencies, energy scale and energy resolution for electrons,
photons and for muons; jet energy scale (JES) and jet energy
resolution, which include a specific treatment for $b$-jets;
contributions to the $E_{\rm T}^{\rm miss}$ uncertainties uncorrelated
with the JES; $b$-tagging and $b$-veto. The effect of these systematic
uncertainties depends on the topology of each final state, but is
typically small compared to that from the theoretical prediction of
the production cross section. The only exception is the jet energy
scale uncertainty which can reach $\sim$20\%\ on the signal yield in
channels such as \hWWlnqq\ and \hZZllqq. The electron and muon energy
scales are directly constrained by $Z\rightarrow e^{+}e^{-}$ and
$Z\rightarrow \mu^{+}\mu^{-}$ events; the impact of the resulting
systematic uncertainty on the four-lepton invariant mass is of the
order of $\sim$0.5\% for electrons and negligible for muons. The
impact of the photon energy scale systematic uncertainty on the
diphoton invariant mass is approximately 0.6\%.

\section{Exclusion Limits}

\begin{figure}[h!]
  \begin{center}
    \includegraphics[width=.49\textwidth]{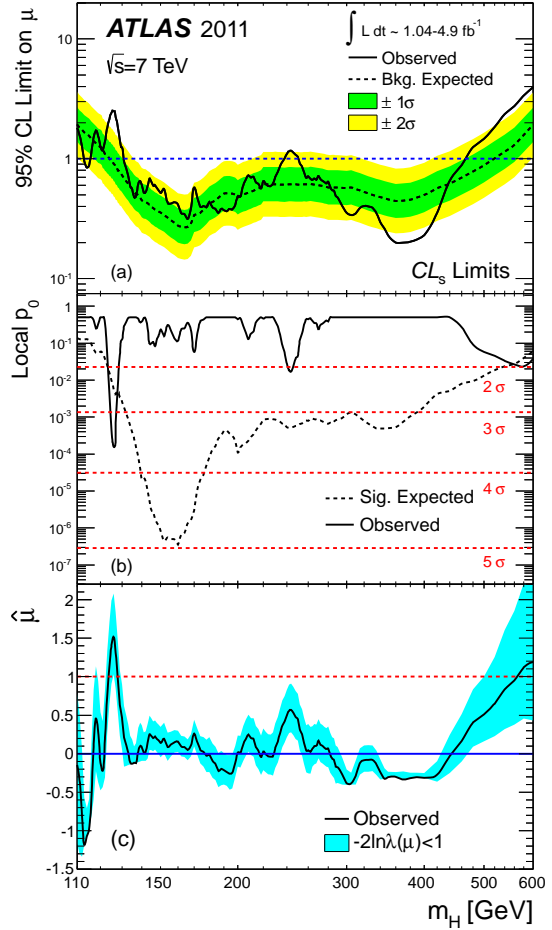}
\vspace{-0.9cm}
  \end{center}
  \caption{(a) The combined 95\%~CL upper limits on the signal
    strength as a function of \mH; the solid curve indicates the
    observed limit and the dotted curve illustrates the median
    expected limit in the absence of a signal together with the $\pm
    1\sigma$ (dark) and $\pm 2\sigma$ (light) bands. (b) The local
    $p_0$ as a function of the \mH\ hypothesis. The dashed curve
    indicates the median expected value for the hypothesis of a SM
    Higgs boson signal at that mass. The four horizontal dashed lines
    indicate the $p_0$ values corresponding to significances of
    2$\sigma$, 3$\sigma$, 4$\sigma$ and 5$\sigma$. (c) The best-fit
    signal strength as a function of the \mH\ hypothesis. The band
    shows the interval around $\hat{\mu}$ corresponding to region
    where $-2\ln \lambda(\mu)<1$.}
  \label{fig:CLs}
\end{figure}

The signal strength, $\mu$, is defined as $\mu=\sigma/\sigma_{\rm
  SM}$, where $\sigma$ is the Higgs boson production cross section
being tested and $\sigma_{SM}$ its SM value; it is a single factor
used to scale all signal production processes for a given \mH\
hypothesis. The combination procedure of
Refs.~\cite{paper2010,LHC-HCG,Moneta:2010pm} is based on the profile
likelihood ratio test statistic $\lambda({\mu})$~\cite{Cowan:2010st},
which extracts the information on the signal strength from the full
likelihood including all the parameters describing the systematic
uncertainties and their correlations. Exclusion limits are based on
the $CL_s$ method~\cite{Read:2002hq} and a value of $\mu$ is regarded
as excluded at the 95\% (99\%)~CL when $CL_s$ takes on the
corresponding value.

The combined 95\%~CL exclusion limits on $\mu$ are shown in
Fig.~\ref{fig:CLs}(a) as a function of \mh. These results are based on
the asymptotic approximation~\cite{Cowan:2010st}.  The observed and
expected limits using this procedure have been validated using
ensemble tests and a Bayesian calculation of the exclusion limits with
a uniform prior on the signal cross section. These approaches agree
with the asymptotic median results to within a few percent.  The
expected 95\%~CL exclusion region covers the \mH\ range from
\lowerExp\ to \upperExp. The observed 95\%~CL exclusion regions are
from \lowerlowerObs\ to \upperlowerObs\ and from \lowerupperObs\ to
\upperupperObs. The regions between 133~\GeV\ and 230~GeV and between
260~\GeV\ and 437~GeV are excluded at the 99\%~CL. A deficit of events
is observed in two \mH\ regions. At very low masses a local deficit in
the diphoton channel allows an additional small mass range between
\lowerIsland\ and \upperIsland\ to be excluded at the 95\%~CL. Small
deficits in various high-mass channels lead to observed limits for
masses between 300~GeV and 400~GeV that are stronger than
expected. The local probability of such a downward fluctuation of a
background-only experiment corresponds to a significance of
approximately 2.5$\sigma$. The probability to observe such a downward
fluctuation over the full inspected mass range in the absence of a
signal, using the method described in
Section~\ref{sec:Significance}~\cite{LEE}, is estimated to be
approximately 30\%.

The observed exclusion covers a large part of the expected exclusion
range, with the exception of the low and high \mH\ regions where
excesses of events above the expected background are observed in
various channels, and in a small mass interval around 245~GeV, which
is not excluded due to an excess mostly in the \hZZllll\ channel.  

\section{Significance of the Excess}
\label{sec:Significance}

An excess of events is observed near \mh$\sim$126~GeV in the \hgg\ and
\hZZllll\ channels, both of which provide a high-resolution invariant
mass for fully reconstructed candidates.  The \hWWlnln\ channel as
well has a broad excess of events in the transverse mass distribution
as seen in Fig.~\ref{fig:lowmass}(d).

The significance of an excess is quantified by the probability ($p_0$)
that a background-only experiment is more signal-like than that
observed. The profile likelihood ratio test statistic is defined such
that $p_0$ cannot exceed
50\%~\cite{Cowan:2010st,LHC-HCG,Moneta:2010pm}.

The local $p_0$ probability is assessed for a fixed \mH\ hypothesis
and the equivalent formulation in terms of number of standard
deviations is referred to as the local significance.  The probability
for a background-only experiment to produce a local significance of
this size or larger anywhere in a given mass region is referred to as
the global $p_0$.  The corresponding reduction in the significance is
referred to as the look-elsewhere effect and is estimated using the
prescription described in Refs.~\cite{LEE,LHC-HCG}.

\begin{figure}[htb]
  \begin{center}
    \includegraphics[width=.42\textwidth]{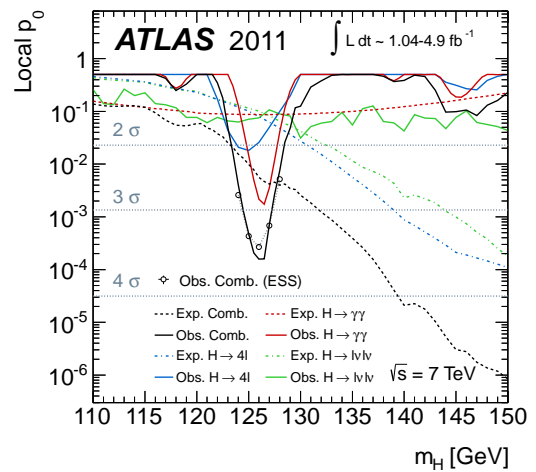} 
     \vspace{-0.7cm}
      \end{center}
      \caption{The local probability $p_0$ for a background-only
        experiment to be more signal-like than the observation. The
        solid curves give the individual and combined observed $p_0$,
        estimated using the asymptotic approximation.  The dashed
        curves show the median expected value for the hypothesis of a
        SM Higgs boson signal at that mass.  The three horizontal
        dashed lines indicate the $p_0$ corresponding to significances
        of 2$\sigma$, 3$\sigma$, and 4$\sigma$. The points indicate
        the observed local $p_0$ estimated using ensemble tests and
        taking into account energy scale systematic uncertainties
        (ESS).}
  \label{fig:p0}
\end{figure}

The observed local $p_0$ values, calculated using the asymptotic
approximation~\cite{Cowan:2010st}, as a function of \mh\ and the
expected value in the presence of a SM Higgs boson signal at that
mass, are shown in Fig.~\ref{fig:CLs}(b) in the entire search mass
range and in Fig.~\ref{fig:p0} for the individual channels and their
combination in the low mass range. Numerically consistent results are
obtained using ensemble tests.

The largest local significance for the combination is achieved for
\mH=126~\GeV, where it reaches \significance\ with an expected value
of 2.5$\sigma$ for a SM signal. The observed (expected) local
significance for \mh$=$126~\GeV\ is 2.8$\sigma$ (1.4$\sigma$) in the
\hgg\ channel, 2.1$\sigma$ (1.4$\sigma$) in the \hZZllll\ channel, and
1.4$\sigma$ (1.4$\sigma$) in the \hWWlnln\ channel.

The significance of the excess is mildly sensitive to systematic
uncertainties on the energy scale (herein referred to as ESS) and
resolution for photons and electrons. The muon energy scale systematic
uncertainties are smaller and therefore neglected. The presence of
these uncertainties, which affect the shape and position of the signal
distributions, lead to a small deviation from the asymptotic
approximation. The observed $p_0$ including these effects is therefore
computed using ensemble tests. The results are displayed in
Fig.~\ref{fig:p0} as a function of \mH.  The observed effect of the
ESS uncertainty is small and reduces the maximum
local significance from 3.6$\sigma$ to 3.5$\sigma$.

The global $p_0$ of a local excess depends on the range of \mH\ and
the channels considered. The global $p_0$ associated with a
2.8$\sigma$ excess anywhere in the \hgg\ search domain 110--150~GeV is
approximately 7\%.  A 2.1$\sigma$ excess anywhere in the \hZZllll\
search range 110--600~GeV corresponds to a global $p_0$ of
approximately 30\%. The global $p_0$ for a combined 3.5$\sigma$ excess
to be found anywhere in the range from 114~GeV to 146~GeV is 0.6\%\
(2.5$\sigma$).  This mass interval corresponds to the region not
excluded at 99\%~CL by the combination of Higgs boson searches at
LEP~\cite{LEP} and the first LHC combined
search~\cite{lhcCombination}. For the full mass range from 110~GeV to
600~GeV, the global $p_0$ is 1.4\%\ (2.2$\sigma$).

The best-fit value of $\mu$, denoted $\hat{\mu}$, is displayed in
Fig.~\ref{fig:CLs}(c) as a function of the \mH\ hypothesis.  The bands
around $\hat{\mu}$ illustrate the $\mu$ interval corresponding to
$-2\ln \lambda(\mu)<1$ and represent an approximate $\pm 1\sigma$
variation.  When evaluating exclusion limits and significance, $\mu$
is not allowed to be negative; however, this restriction is not
applied in Fig.~\ref{fig:CLs}(c), in order to illustrate the presence
and extent of downward fluctuations.  Nevertheless, the $\mu$
parameter is still bounded to prevent negative values of the
probability density functions in the individual channels, and for
negative $\hat{\mu}$ values close to the boundary, the $-2\ln
\lambda(\mu)<1$ region does not always reflect a 68\%\ confidence
interval. The excess observed for $\mH=126\GeV$ corresponds to
$\hat{\mu}$ of approximately $1.5^{+0.6}_{-0.5}$, which is compatible
with the signal expected from a SM Higgs boson at that mass ($\mu=1$).

\section{Conclusions}

A dataset of up to 4.9~\infb\ recorded in 2011 has been used to search
for the SM Higgs boson with the ATLAS experiment at the LHC.  Higgs
boson masses between \lowerExp\ and \upperExp\ are expected to be
excluded at the 95\%~CL. The observed exclusion at the 95\%~CL ranges
from \lowerIsland\ to \upperIsland, \lowerlowerObs\ to \upperlowerObs\
and \lowerupperObs\ to \upperupperObs. An exclusion of the SM Higgs
boson production at the 99\%~CL is achieved in the regions between
133~GeV and 230~GeV and between 260~GeV and 437~GeV.

An excess of events is observed in the \hgg\ and \hZZllll\ channels,
for \mH\ close to 126~GeV, which is also supported by a broad excess
in the \hWWlnln\ channel. The observed local significances of the
individual excesses are 2.8$\sigma$, 2.0$\sigma$ and 1.4$\sigma$,
respectively. The expected local significances of these channels, for
a 126~GeV SM Higgs boson are, coincidentally, all
$\sim$1.4$\sigma$. The combined local significance of these excesses
is \significance. When the energy scale uncertainties are taken into
account, the combined local significance is reduced to
\significanceESS. The expected combined local significance in the
presence of a SM Higgs boson signal at that mass is 2.5$\sigma$.  The
global probability for such an excess to be found in the full search
range, in the absence of a signal, is approximately 1.4\%,
corresponding to 2.2$\sigma$.

\section{Acknowledgments}

  We thank CERN for the very successful operation of the LHC, as well
  as the support staff from our institutions without whom ATLAS could
  not be operated efficiently.  

  We acknowledge the support of ANPCyT, Argentina; YerPhI, Armenia;
  ARC, Australia; BMWF, Austria; ANAS, Azerbaijan; SSTC, Belarus; CNPq
  and FAPESP, Brazil; NSERC, NRC and CFI, Canada; CERN; CONICYT,
  Chile; CAS, MOST and NSFC, China; COLCIENCIAS, Colombia; MSMT CR,
  MPO CR and VSC CR, Czech Republic; DNRF, DNSRC and Lundbeck
  Foundation, Denmark; ARTEMIS and ERC, European Union; 
  IN2P3-CNRS, CEA-DSM/IRFU, France; GNAS, Georgia; BMBF, DFG, HGF, MPG
  and AvH Foundation, Germany; GSRT, Greece; ISF, MINERVA, GIF, DIP
  and Benoziyo Center, Israel; INFN, Italy; MEXT and JSPS, Japan;
  CNRST, Morocco; FOM and NWO, Netherlands; RCN, Norway; MNiSW,
  Poland; GRICES and FCT, Portugal; MERYS (MECTS), Romania; MES of
  Russia and ROSATOM, Russian Federation; JINR; MSTD, Serbia; MSSR,
  Slovakia; ARRS and MVZT, Slovenia; DST/NRF, South Africa; MICINN,
  Spain; SRC and Wallenberg Foundation, Sweden; SER, SNSF and Cantons
  of Bern and Geneva, Switzerland; NSC, Taiwan; TAEK, Turkey; STFC,
  the Royal Society and Leverhulme Trust, United Kingdom; DOE and NSF,
  United States of America.

  The crucial computing support from all WLCG partners is acknowledged
  gratefully, in particular from CERN and the ATLAS Tier-1 facilities
  at TRIUMF (Canada), NDGF (Denmark, Norway, Sweden), CC-IN2P3
  (France), KIT/GridKA (Germany), INFN-CNAF (Italy), NL-T1
  (Netherlands), PIC (Spain), ASGC (Taiwan), RAL (UK) and BNL (USA)
  and in the Tier-2 facilities worldwide.

\bibliographystyle{elsarticle-num} 
\bibliography{paper}

\clearpage 
\onecolumn
\begin{flushleft}
{\Large The ATLAS Collaboration}

\bigskip

G.~Aad$^{\rm 48}$,
B.~Abbott$^{\rm 110}$,
J.~Abdallah$^{\rm 11}$,
S.~Abdel~Khalek$^{\rm 114}$,
A.A.~Abdelalim$^{\rm 49}$,
A.~Abdesselam$^{\rm 117}$,
O.~Abdinov$^{\rm 10}$,
B.~Abi$^{\rm 111}$,
M.~Abolins$^{\rm 87}$,
O.S.~AbouZeid$^{\rm 157}$
% ...
\bigskip

% ...
$^{10}$ Institute of Physics, Azerbaijan Academy of Sciences, Baku, Azerbaijan\\
$^{11}$ Institut de F\'isica d'Altes Energies and Departament de F\'isica de la Universitat Aut\`onoma  de Barcelona and ICREA, Barcelona, Spain\\
% ...
$^{48}$ Fakult\"{a}t f\"{u}r Mathematik und Physik, Albert-Ludwigs-Universit\"{a}t, Freiburg i.Br., Germany\\
$^{49}$ Section de Physique, Universit\'e de Gen\`eve, Geneva, Switzerland\\
% ...
$^{87}$ Department of Physics and Astronomy, Michigan State University, East Lansing MI, United States of America\\
% ...
$^{110}$ Homer L. Dodge Department of Physics and Astronomy, University of Oklahoma, Norman OK, United States of America\\
$^{111}$ Department of Physics, Oklahoma State University, Stillwater OK, United States of America\\
% ...
$^{114}$ LAL, Univ. Paris-Sud and CNRS/IN2P3, Orsay, France\\
% ...
$^{117}$ Department of Physics, Oxford University, Oxford, United Kingdom\\
% ...
$^{157}$ Department of Physics, University of Toronto, Toronto ON, Canada\\
% ...
\end{flushleft}


\end{document}